\documentclass[aps,prl,preprint]{revtex4}

\usepackage{graphicx}
\usepackage[export]{adjustbox}
\usepackage{amsmath}
\usepackage{float}

\begin{document}

\date{\today}
\newcommand{\Ca}{$^{40}$Ca$^+$}
\newcommand{\CaH}{$^{40}$CaH$^+$}



%
\title{Preparation and coherent manipulation of pure quantum states of a single molecular ion}
%

\author{Chin-wen Chou$^1$, Christoph Kurz$^{1,2}$, David B. Hume$^1$, Philipp~N.~Plessow$^3$, David R. Leibrandt$^{1,2}$, and Dietrich Leibfried$^1$}

\affiliation{$^1$Time and Frequency Division, National Institute of Standards and Technology, Boulder, Colorado 80305, USA}
\affiliation{$^2$University of Colorado, Boulder, Colorado, USA}
\affiliation{$^3$Institute of Catalysis Research and Technology, Karlsruhe Institute of Technology, Karlsruhe, Germany}
\maketitle

\textbf{Laser cooling and trapping of atoms and atomic ions has led to numerous advances including the observation of exotic phases of matter~\cite{bloch08,ioncrystal}, development of exquisite sensors~\cite{kitching11} and state-of-the-art atomic clocks~\cite{ludlow15}. The same level of control in molecules could also lead to profound developments such as controlled chemical reactions and sensitive probes of fundamental theories~\cite{schiller05}, but the vibrational and rotational degrees of freedom in molecules pose a formidable challenge for controlling their quantum mechanical states.  Here, we use quantum-logic spectroscopy (QLS)~\cite{schmidt05} for preparation and nondestructive detection of quantum mechanical states in molecular ions~\cite{wolf16}.  We develop a general technique to enable optical pumping and preparation of the molecule into a pure initial state. This allows for the observation of high-resolution spectra in a single ion (here CaH$^+$) and coherent phenomena such as Rabi flopping and Ramsey fringes.  The protocol requires a single, far-off resonant laser, which is not specific to the molecule, so that many other molecular ions, including polyatomic species, could be treated with the same methods in the same apparatus by changing the molecular source. Combined with long interrogation times afforded by ion traps, a broad range of molecular ions could be studied with unprecedented control and precision, representing a critical step towards proposed applications, such as precision molecular spectroscopy, stringent tests of fundamental physics, quantum computing, and precision control of molecular dynamics~\cite{carr09}.}

Significant progress has been made in recent years toward the goals of controlling the quantum mechanical states of ultracold molecules~\cite{shuman10,ospelkaus10} (also see Methods).  For a molecular ion, its charge provides a means of trapping and sympathetically cooling via its Coulomb interaction with a co-trapped atomic ion that is readily laser-cooled~\cite{barrett03}. Cooling of vibrational~\cite{rellergert13} and rotational~\cite{staanum10,schneider10,hansen14,lien14} states has also been realized in heteronuclear molecular ions.  Preparation in specific vibrational and rotational states was achieved via threshold photoionization~\cite{tong10} and optical pumping into individual hyperfine states has been demonstrated~\cite{bressel12}. In the context of QLS, state detection of a single molecular ion in a particular subset of states in a rotational manifold has been achieved~\cite{wolf16}. Many of these experiments rely on fortuitous molecular properties~\cite{shuman10,lien14}, dedicated multi-laser systems~\cite{shuman10,staanum10,schneider10,bressel12} or sophisticated laser cooling techniques~\cite{ospelkaus10}. Coherent control of pure quantum states of a molecular ion, crucial to precision experiments, has not yet been accomplished.

Here, we demonstrate a general protocol for coherent manipulation of trapped molecular ions in their electronic and vibrational ground states based on QLS~\cite{schmidt05} and stimulated Raman transitions (SRTs) driven by a far-detuned laser source \cite{schmidt06,leibfried12,ding12}. Because the rotational motion is not cooled, our approach relies on probabilistically preparing a particular rotational state via a projective measurement~\cite{vogelius06}. We cool the shared motion of the molecular ion and a co-trapped atomic ion to the ground state~\cite{barrett03}. Then we set the relative detuning of the Raman beams to drive a specific transition in the molecule in such a way that a state change of the molecule is accompanied by an excitation of the shared motion (motional sideband). We can efficiently detect this excitation with the atomic ion, which projects the molecule into the final state of the transition, leaving the molecule in a known, pure quantum state~\cite{vogelius06}. This allows for subsequent manipulation of the molecular state, as well as spectroscopy of molecular transitions. In addition, we pump the molecular ion into specific sublevels of its rotational states, effectively orienting the molecular rotation along an axis of our choice. The improved orientation after pumping increases the state preparation success rate and the signal-to-noise ratio in subsequent experiments. 

\begin{figure}
\includegraphics[angle=0, width=0.8\textwidth]{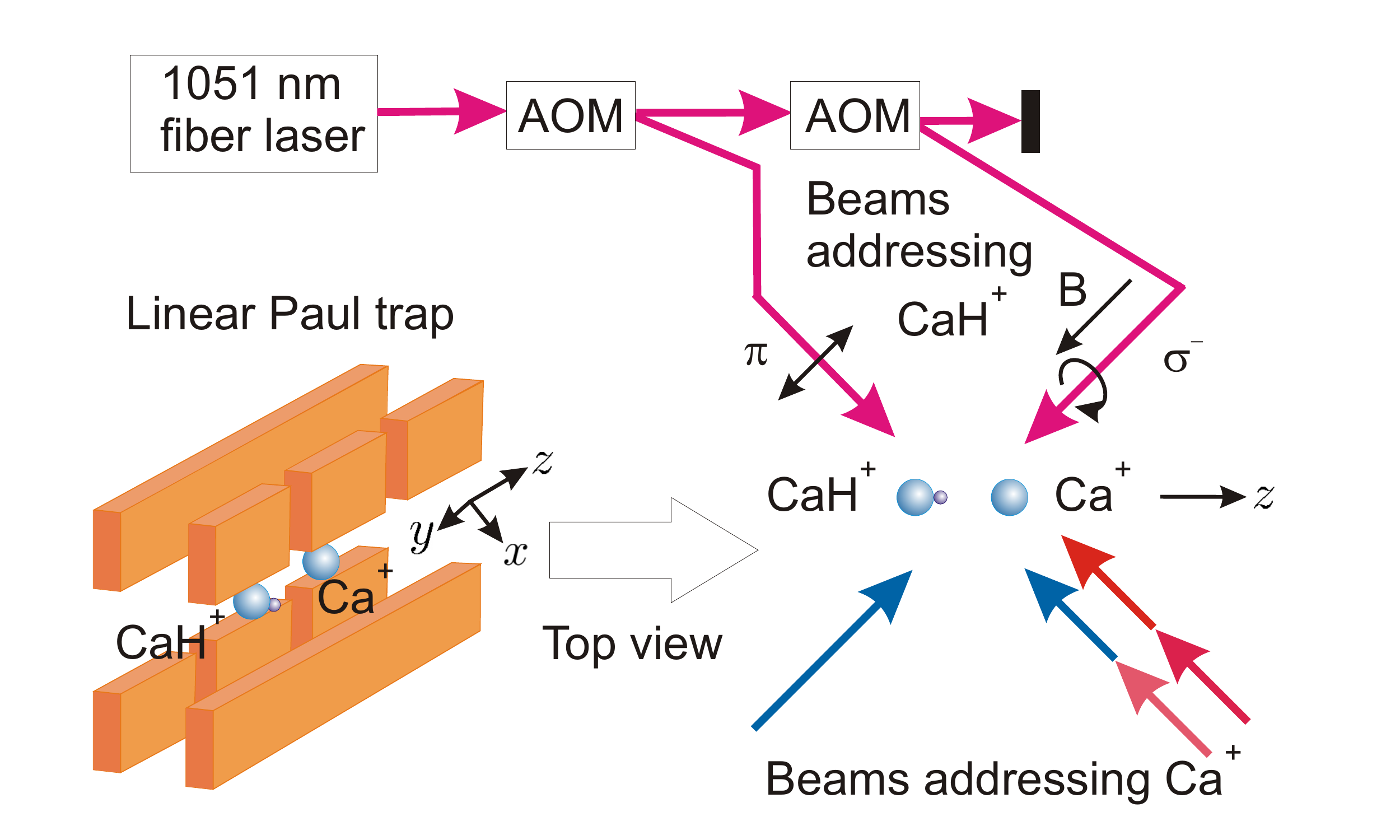}
\caption[Apparatus]{Simplified experimental setup. A \CaH-\Ca~ion pair is held in a linear Paul trap. Two modes of motion in the $z$ direction and one in the $x$ direction of the ion pair are prepared in the motional ground states via laser cooling on \Ca. The motional mode in the z direction in which the two ions oscillate out of phase is used as the normal mode for the QLS protocol. Two Raman beams derived from a single fiber laser are split in frequency by two acousto-optic modulators (AOMs). They are directed onto the molecular ion with a $\mathbf{k}$-vector difference along the trap $z$ axis. The two Raman beams have $\pi$ and $\sigma^-$ polarizations relative to the quantization axis defined by the applied magnetic field. They drive either carrier two-photon stimulated-Raman transitions in the molecule, or their motional sidebands, depending on the frequency detuning, while changing the projection quantum number $m$ of the molecular angular momentum by $\pm1$. Single quanta of excitation in the ion motion can be detected by driving sidebands of the narrow quadrupole transition between the 4s$^2$S$_{1/2}$ and 3d$^2$D$_{5/2}$ levels of the \Ca~atomic ion, followed by electron shelving detection with the laser beams addressing \Ca~(see text and Methods). Detection of motional excitation on the \Ca~ion projects the molecule into the final state of the addressed transition, which is then available for further manipulation.}
\label{Fig1}
\end{figure}

In our experiments, we trap two \Ca~ions  in  a harmonic ion trap in ultra-high vacuum at room temperature ($P\sim4\times10^{-9}$~Pa). To form the molecular ion, hydrogen gas is leaked into the vacuum chamber until one of the \Ca~ions reacts to form a \CaH~ion, which quickly relaxes to its singlet electronic and vibrational ground state, but remains in a mixture of rotational states, in equilibrium with the blackbody radiation of its room temperature environment~\cite{kimura11}. The \CaH~molecular ion serves as a test case for a much wider class of molecules that could be generated by various other techniques (see Methods). The \Ca~ion, which can be readily laser cooled, optically pumped, and manipulated \cite{roos99}, is coupled to the molecule by mutual Coulomb repulsion so that the shared normal modes of translational motion can be sympathetically cooled to their ground states \cite{barrett03, wolf16}.  A simplified diagram of the experimental setup is shown in Fig.~\ref{Fig1}.

%

\begin{figure}
\includegraphics[angle=0, width=0.50\textwidth]{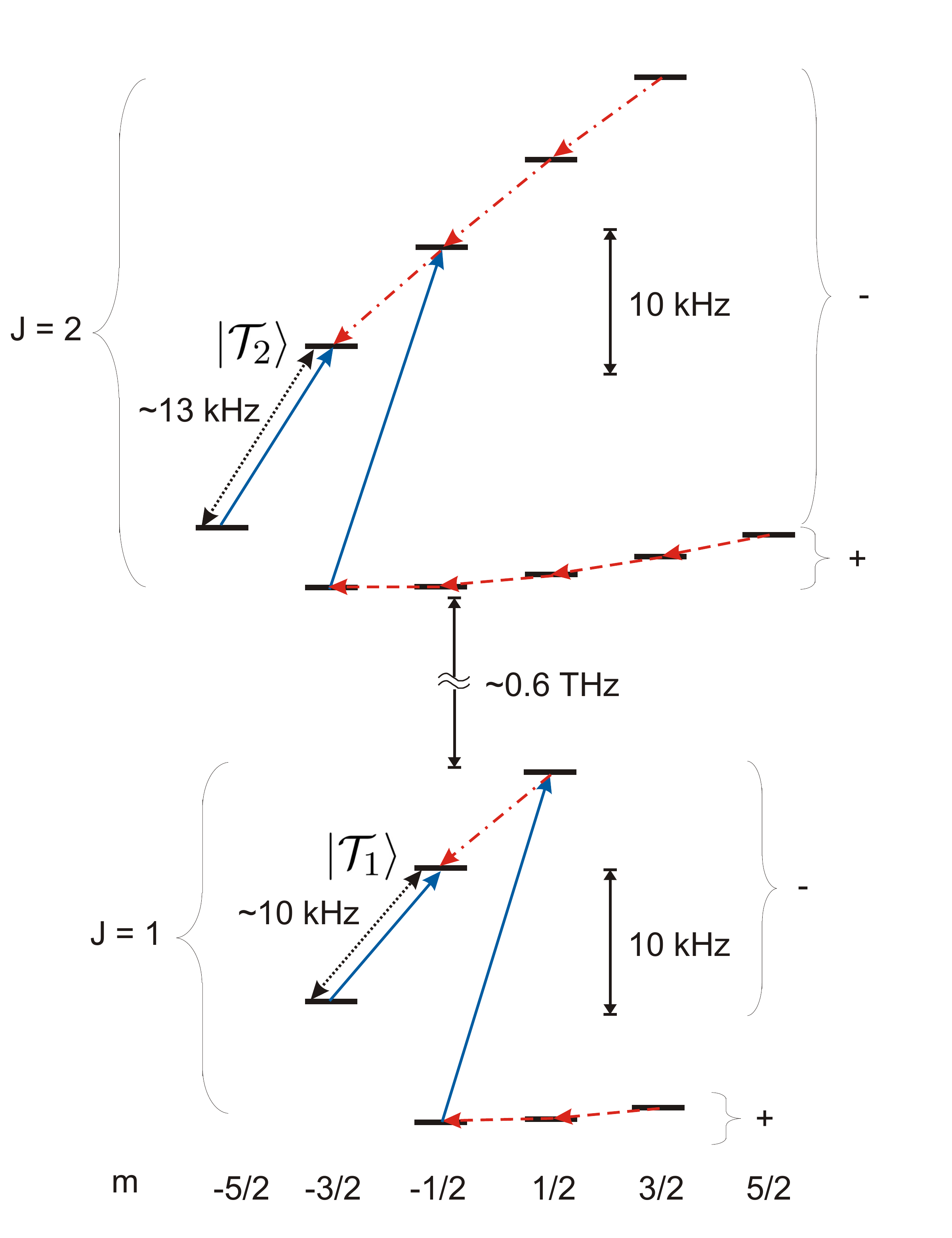}
\caption[Level diagram]{The level diagrams for the rotational levels of \CaH~for $J\in\{1,2\}$. At $|\mathbf{B}|\approx0.36$~mT, the energy eigenstates $|J,m, \pm\rangle$ are either superpositions of product states of rotation and nuclear spin sublevels adding to the same total $m$, or extreme states where the proton spin and the rotation are aligned ($m=\pm(J+1/2)$). The red dashed, red dot-dashed, and blue solid arrows indicate the directionality of the optical pumping transitions for population concentration. Red dashed and dot-dashed arrows indicate that several transitions can be addressed simultaneously within the spectral resolution of the pumping sequence, while blue arrows indicate transitions with resolved frequencies. The black dotted arrows show the $|\mathcal{T}_J\rangle=|J,-J+1/2,-\rangle\leftrightarrow|J,-J-1/2,-\rangle$ target transitions with unique frequencies that are driven for the projective pure state preparation and coherent manipulation (see text).} 
\label{Fig2}
\end{figure}

 Rotation of the molecule and coupling of rotational angular momentum $\hat{\mathbf{J}}$, spin of the proton $\hat{\mathbf{I}}$, and external magnetic field $\mathbf{B}$, is described by the Hamiltonian:
 \begin{equation}\label{Hamiltonian}
   \hat{H}_{rot} = \frac{1}{\hbar} \bigg( 2\pi R \hat{\mathbf{J}}^2 -g \mu_{N}\hat{\mathbf{J}}\cdot \mathbf{B}-g_I \mu_{N}\hat{\mathbf{I}}\cdot \mathbf{B}- 2\pi c_{IJ} \hat{\mathbf{I}}\cdot \hat{\mathbf{J}}\bigg)\text{,}
 \end{equation}
  where  $R\approx 144$~GHz is the rotational constant for \CaH~\cite{abe2012}, $\mu_{N}$ is the nuclear magneton, $g$ and $g_I$ are the g-factors, and $c_{IJ}$ is the spin-rotation constant (see Methods). With $|\mathbf{B}|\approx$~0.36~mT, we can calculate energy levels of the eigenstates of $\hat{H}_{rot}$. Those for rotational quantum number $J \in \{1,2\}$ are displayed in Fig.~\ref{Fig2}. We classify the eigenstates by $|\mathcal{J}\rangle\equiv|J,m, \xi\rangle$, where $\mathcal{J}$ stands for the set of quantum numbers $\{J,m, \xi\}$ and $m\in\{-J-1/2,-J+1/2...,J+1/2\}$ denotes the sum $m=m_J+m_I$ of the components of the rotational angular momentum and the proton spin along $\mathbf{B}$. The value of $m$ is a good quantum number for arbitrary $\mathbf{B}$. The last label $\xi\in\{+,-\}$ indicates the relative sign in the superposition of product states with the same $m$ but opposite proton spin, $|J,m_J=m+1/2\rangle|m_I=-1/2\rangle$ and $|J,m_J=m-1/2\rangle|m_I=+1/2\rangle$ (see Methods). In the extreme states where spin and rotational angular momentum are aligned with the quantization axis, the eigenstates are simple product states
$|J,\pm (J+1/2), \pm\rangle= |J, m_J=\pm J\rangle |m_I=\pm 1/2\rangle$ and the label $\xi$ denotes the sign of $\pm (J+1/2)$. We operate at an intermediate magnetic field where the last three coupling terms in Eq.~(\ref{Hamiltonian}) lead to energy shifts of similar magnitude.

To drive the SRTs in the molecule, we use two Raman beams from a continuous-wave fiber laser with wavelength $\approx 1051$~nm. The two beams are directed separately onto the ions, one circularly polarized ($\sigma^-$), the other linearly ($\pi$) with respect to the quantization axis along the magnetic field, as indicated in  Fig.~\ref{Fig1}. With $\Delta\nu\equiv\nu_\sigma-\nu_\pi$ tuned near $\pm(E_f-E_i)/h$ and $\pm[(E_f-E_i)/h\pm\nu_t]$, they can coherently drive carrier two-photon SRTs and the first sidebands of translational motion, respectively. Here $\nu_\sigma$ ($\nu_\pi$) is the frequency of the $\sigma^-$ ($\pi$) polarized beam, $E_i$ ($E_f$) is the energy of the initial (final) molecular level, and $\nu_t$ is the frequency of the motional mode. To minimize the perturbation on the molecular energy levels caused by the Raman beams, we control the intensity ratio between the $\sigma^-$ and $\pi$ beams to minimize the differential AC Stark shifts between the molecular levels (see Methods).

\begin{figure}
\vspace{-3.0 in}
\includegraphics[angle=0, width=1.0\textwidth]{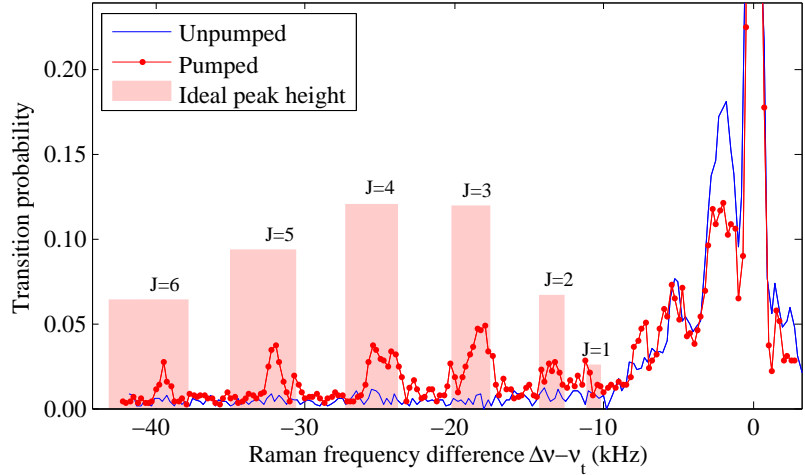}
\vspace{-3 in}
\caption[Raman Spectra]{Raman spectra for $\Delta m = -1$ blue sideband transitions probed with 1 ms pulses. The frequency axis shows the offset from the motional frequency of the normal mode at $\nu_t\approx5.164$~MHz. Data without optical pumping (blue line) shows two prominent peaks near $-2$~kHz and $-6$~kHz that correspond to the transitions $|J,m,\xi\rangle\rightarrow|J,m-1,\xi\rangle$ which overlap for different m, for the two cases $\xi\in\{+,-\}$, shown by the red dashed and dot-dashed arrows, respectively, in Fig.~\ref{Fig2}. Peaks between $0$ and $-10$ kHz are visible because they arise from a large number of sublevels. The peak at 0~kHz is attributed to coherent motion that can be driven by modulated optical dipole forces without changing the internal state of the molecule if the polarization of the Raman beams is not perfectly orthogonal on both ions. Data with optical pumping (red dots) shows peaks near $\{-11$, $-14$, $-19$, $-25$, $-32$, $-40\}$~kHz, which are identified as the $|J,-J+\frac{1}{2},-\rangle\rightarrow|J,-J-\frac{1}{2},-\rangle$ transitions for $J\in\{1, 2, 3, 4, 5, 6\}$, respectively, from their proximity with the predicted frequencies (see Methods). Pink-shaded bars: the frequencies and heights of the peaks predicted by the theory, with the effect from $J$-dependent Rabi rates taken into account. The widths of the bars indicate the ranges of the predicted transition frequencies (see Methods).}
\label{Fig3}
\end{figure}

Based on the framework of QLS, the probabilistic projective preparation of a pure quantum state in the molecular ion proceeds as follows.  If we assume perfect ground state cooling, with the \Ca~prepared in $|D\rangle\equiv|D_{5/2},m=-5/2\rangle$, the density matrix of the molecule, the normal mode, and the atomic ion can be written as
\begin{equation}\label{DMinit}
  \rho_0 = \left( \sum_{\mathcal{J}} P_{\mathcal{J}} |\mathcal{J}\rangle \langle \mathcal{J}|\right)~|0\rangle \langle0|~
  |D\rangle \langle D|\text{, }
\end{equation}
where $P_{\mathcal{J}}$ is the population in state $|\mathcal{J}\rangle$, and $|n\rangle$ denotes the motional state with $n$ phonons in the normal mode. With probability $P_{\mathcal{J}_i}$, the molecule is in $|\mathcal{J}_i\rangle$ and we can drive the molecular blue sideband transition $|\mathcal{J}_i\rangle |0\rangle |D\rangle \rightarrow (\alpha|\mathcal{J}_i\rangle |0\rangle +\beta|\mathcal{J}_f\rangle |1\rangle)|D\rangle$, with $|\alpha|^2+|\beta|^2=1$ and selection rule $m_f=m_i\pm1$. If the $|\mathcal{J}_i\rangle\leftrightarrow|\mathcal{J}_f\rangle$ transition has a unique frequency in the molecule, the density matrix of the system is modified to
\begin{eqnarray}\label{DMbsb}
   \rho_1& =&P_{\mathcal{J}_i}~(\alpha|\mathcal{J}_i\rangle |0\rangle +\beta|\mathcal{J}_f\rangle |1\rangle)~(\alpha^*\langle\mathcal{J}_i| \langle 0| +\beta^*\langle \mathcal{J}_f|\langle 1|)~|D\rangle \langle D| \nonumber\\
  &&+\sum_{\mathcal{J} \neq \mathcal{J}_i} P_{\mathcal{J}} |\mathcal{J}\rangle \langle \mathcal{J}|
  ~|0\rangle \langle0|~|D\rangle \langle D|\text{.}
\end{eqnarray}
At this point we can attempt to drive a $\pi$-pulse on the red sideband of \Ca, which induces the transition $|1\rangle |D\rangle \rightarrow |0\rangle |S\rangle$, where $|S\rangle$ denotes the $S_{1/2}$, $m=-1/2$ state of \Ca. Fluorescence detection on the \Ca~distinguishes $|S\rangle$, which scatters many photons, from $|D\rangle$, which ideally scatters no photons. High detection fidelity can be achieved in a single shot by simply checking the resultant photon counts against a pre-determined threshold (see Methods and~\cite{leibfried03,myerson08}). Thus, with probability $P_{\mathcal{J}_i}|\beta|^2$, the density matrix of the molecule is projected to the pure state
\begin{equation}\label{DMproj}
  \rho_M =  |\mathcal{J}_f\rangle \langle \mathcal{J}_f|\text{,}
\end{equation}
which is heralded non-destructively by detecting that the \Ca~ion scatters photons. We can measure Raman spectra of molecular transitions by recording the probability of detecting $|S\rangle$, as long as we leave the molecule enough time to re-establish equilibrium with the blackbody environment, so that $P_{\mathcal{J}_i}$ is the same for every attempt. Such a spectrum with $m_f =m_i -1$ is shown as the blue curve in Fig.~\ref{Fig3}. Due to the small magnitudes of $P_{\mathcal{J}}$ ($<1.1$~\% at 300~K), the transitions with frequencies $\omega/2\pi\equiv(E_f-E_i)/h$ between $-10$ and $-50$ kHz are not discernible from measurement noise, which we have optimized to be at the 0.5~\% level (see Methods).

In order to achieve a better signal-to-noise ratio in the target transitions with unique frequencies ($|J,-J+1/2,-\rangle\leftrightarrow|J,-J-1/2,-\rangle$ in this work), it is beneficial to increase the populations in the corresponding initial states over the thermal equilibrium levels. Since the blackbody relaxation times ($\sim$100~ms to $>2$~s at 300 K for $J<8$) are long compared with attempts to drive transitions ($<5$~ms), we can concentrate the population via optical pumping. The pumping uses a pulse sequence similar to projective state preparation.  The ions are first cooled to the ground state of motion, then a blue sideband pulse is applied to the molecular transition to be pumped. Further ground state cooling produces dissipation in the system, making these pumping transitions directional~\cite{schmidt05}, analogous to a typical optical pumping process where spontaneous decay removes entropy from the system.
The effect of pumping in the molecular system can be understood by returning to Eq.~(\ref{DMbsb}), which describes the density matrix of the system after the blue sideband pumping pulse on the molecular ion.  In this case $J_i$  denotes the state being pumped.  If $|\beta| \neq 0$ after the pumping attempt, the sideband cooling pulses on the \Ca~destroy the coherence induced from driving the molecular blue sideband transition, transforming the density matrix from $\rho_1$ to
\begin{equation}\label{DMcool}
  \rho_2 = \left[P_{\mathcal{J}_i}~|\alpha|^2 |\mathcal{J}_i\rangle \langle\mathcal{J}_i| +(P_{\mathcal{J}_i} |\beta|^2+P_{\mathcal{J}_f}) |\mathcal{J}_f\rangle\langle \mathcal{J}_f| +
 \sum_{\mathcal{J} \neq \mathcal{J}_i,\mathcal{J}_f} P_{\mathcal{J}} |\mathcal{J}\rangle \langle \mathcal{J}|
 ~\right]~|0\rangle \langle0|~|D\rangle \langle D|\text{.}
\end{equation}
The population of $|\mathcal{J}_i\rangle$ is decreased to $P_{\mathcal{J}_i}|\alpha|^2$ while the population of $|\mathcal{J}_f\rangle$ is increased to $P_{\mathcal{J}_f}+P_{\mathcal{J}_i}|\beta|^2$. When repeating this sequence on the transitions indicated by the blue solid, red dashed, and red dot-dashed arrows in Fig.~\ref{Fig2}, the directionality of angular momentum transfer pumps the population toward the target states, which for each $J\in\{1,2,3,...\}$ are denoted by the set of quantum numbers $\mathcal{T}_J=\{J,m = -J+\frac{1}{2},-\}$. We could pump to a different subset of states by modifying the choice of transitions and their directionality. The quantization axis and beam polarizations can be chosen to orient the rotational axis as well as the proton spin of the resultant pumped molecular state as desired.

For \CaH~in $|\mathbf{B}|=0.36$~mT, the frequencies of the transitions $|J, m, +\rangle\rightarrow|J, m-1, +\rangle$ and $|J, m, -\rangle\rightarrow|J, m-1, -\rangle$, indicated by the red dashed and dot-dashed arrows in Fig~\ref{Fig2}, are in two narrow regions around $-2$~kHz and $-6$~kHz, respectively (see Fig.~\ref{Fig2} and Fig.~\ref{Fig3}), thus multiple transitions involving significant population can be simultaneously pumped. In order to pump the majority of the population to the target states, transitions with $\omega/2\pi>10$ kHz (blue solid arrows in Fig~\ref{Fig2}) also need to be addressed with additional pumping pulses. Assuming perfect pumping efficiency, the steady state density matrix can be approximated by
\begin{equation}\label{DMpump}
  \rho_p = \bigg(\frac{1}{2}Q_0\sum_{J=0}|\mathcal{J}\rangle\langle\mathcal{J}|+\sum_{J>0} Q_{J} |J, -J+\frac{1}{2},-\rangle \langle J, -J+\frac{1}{2},-|\bigg)~|0\rangle \langle0|~|D\rangle \langle D|\text{, }
\end{equation}
where $Q_{J}=\sum_{m,\xi}P_{J,m,\xi}$ is the total probability of being in the manifold with rotational quantum number $J$, which is increased by a factor $2(2 J+1)$ (the number of sublevels in the manifold) over the probability of being in a certain sublevel $|J,m,\xi\rangle$ in the equilibrium distribution. The Raman spectrum taken after the pumping stage is shown as the linked red dots in Fig.~\ref{Fig3}. The excitation probabilities for the isolated target transitions are significantly increased over the equilibrium spectrum, but we infer a pumping efficiency $<50~\%$.

After optical pumping, the frequencies of the target transitions are distinguishable and allow for projective state preparation. False-positive detection of changes in phonon number can be due to imperfect ground state cooling or heating of the motion from other sources, which impacts the fidelity of the projective state preparation. We can improve the preparation fidelity by successively driving the $|\mathcal{T}_J\rangle|0\rangle\leftrightarrow|J,-J-\frac{1}{2},-\rangle|1\rangle$ sideband transition with $\pi$ pulses, interspersed by monitoring the creation and removal of phonons of motion and ground state cooling. Repeated detection of phonon number changes increases the preparation fidelity of the desired state to approximately 80~\% (see Methods). 

\begin{figure}
\vspace{-1.5 in}
\includegraphics[angle=0, width=0.95\textwidth]{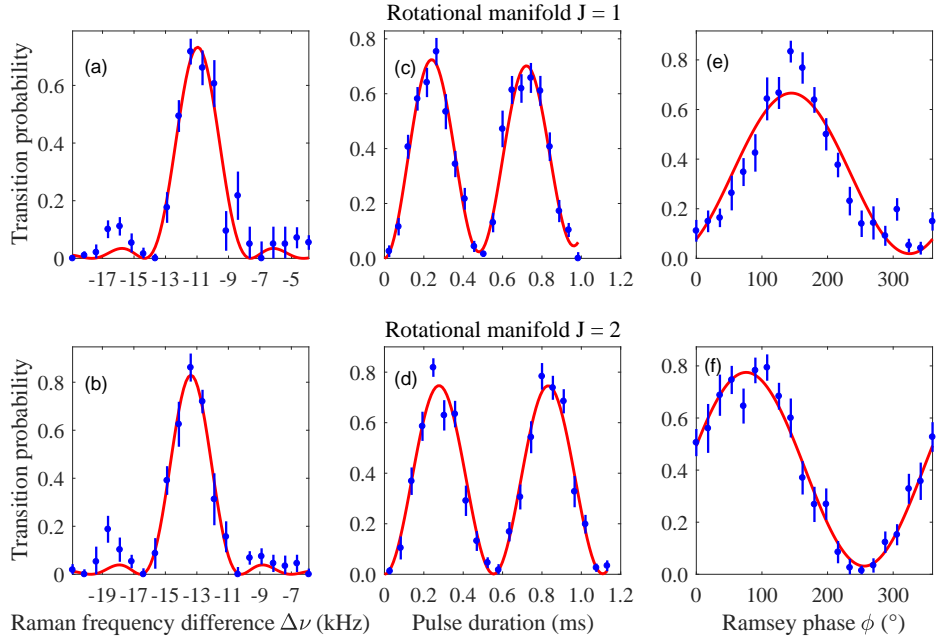}
\vspace{-2.0 in}
\caption[Rabi flopping]{Coherent spectroscopy and manipulation of pure molecular states. The $|J, -J-1/2, -\rangle$ state with $J \in \{1,2\}$ is prepared with adaptive molecular state preparation, followed by interrogation of the $|J,-J-\frac{1}{2},-\rangle\leftrightarrow|\mathcal{T}_J\rangle=|J,-J+\frac{1}{2},-\rangle$ transition. (a) and (b): Frequency spectra; (c) and (d): Rabi flopping. The red lines in (a) and (b) ((c) and (d)) are the fits to sinc-square functions (exponentially damped sinusoids). (e) and (f): Ramsey fringes as a function of the relative phase $\phi$ between the two $\pi/2$ pulses with a wait time of 15 ms for (e) $J = 1$ and (f) $J = 2$. The red lines are fits to sinusoidal functions with periods of 360$^\circ$. The error bars stand for $\pm$ s.d. Data acquisition time: approximately 30 minutes for (a) and (b), 1 hour for (c) and (d), and 45 minutes for (e) and (f).}
\label{FigRabi}
\end{figure}

After state preparation, the resultant quantum state can be coherently manipulated. We observe Rabi flopping of the molecule by driving the carrier transition $|\mathcal{T}_J\rangle\leftrightarrow|J,-J-\frac{1}{2},-\rangle$ for different durations and detecting changes in the molecular state by checking whether $\pi$ pulses on the sideband transition $|\mathcal{T}_J\rangle|0\rangle\leftrightarrow|J,-J-\frac{1}{2},-\rangle|1\rangle$ alter the motional state. To avoid driving multiple transitions, the carrier Raman pulses are reduced in power by a factor of 10 compared to the sideband pulses. During the coherent evolution, the state of the molecule is ideally
\begin{equation}
|\Psi(t)\rangle =\cos(\Omega_J t)|J,-J-\frac{1}{2},-\rangle+ \sin(\Omega_J t)|\mathcal{T}_J\rangle,
\end{equation}
where $\Omega_J$ is the $J$-dependent Rabi frequency of the transition (see Methods).
The molecule is detected in $|\mathcal{T}_J\rangle$ with probability $C_J \sin^2(\Omega_J t)$ where $C_J<1$ because of inefficiencies in preparation and detection and imperfections in the driving pulse. By repeating preparation, coherent evolution, and state detection we build up sufficient statistics to observe the Rabi flops shown in Fig.~\ref{FigRabi} as a function of Raman pulse durations for $J\in\{1,2\}$.

To further characterize the coherence, we implement a Ramsey sequence. We first prepare an equally weighted superposition $(|\mathcal{T}_J\rangle+|J,-J-\frac{1}{2},-\rangle)/\sqrt{2}$ by choosing $\Delta\nu=(E_f-E_i)/h$ and $\Omega_J t=\pi/4$ to apply a carrier $\pi/2$ pulse. We then wait for a duration $T$ and apply another $\pi/2$ pulse with a variable phase $\phi$ relative to the first pulse. If the molecular state remains coherent through the second interaction, the probability to return to $|\mathcal{T}_J\rangle$ varies sinusoidally with the relative phase. Fig.~\ref{FigRabi}(e) and (f) show this for $J\in\{1,2\}$ and $T=$~15 ms. The fringes are suitable for metrology with sub-100 Hz resolution, although the systematic uncertainties have not yet been characterized at this level. The stability of the Raman beams could be further improved to better exploit the lifetime of the molecule in the rotational sublevels and achieve higher spectroscopic resolution with a longer Ramsey sequence.

 The main criteria for applying the protocol demonstrated in this work to other molecular ion species are: a. efficient trapping and sympathetic cooling via an atomic ion; b. transitions in the molecule with accessible frequencies and sufficient coherence for probabilistic projective preparation and coherent manipulation of pure states (see Methods). The same laser and methods in our setup are readily applicable to precision spectroscopy and quantum state manipulation of a wide range of molecular species, including symmetric molecules such as H$_2^+$~\cite{ubachs16, schiller14, karr14}, and possibly molecules that have been identified as candidates to impose stringent limits on the electric dipole moment of the electron~\cite{meyer06,loh13}, or the time dependence of fundamental constants (see Methods and references therein). In future work, we plan to use a frequency comb as a second source of Raman pulses that can span the energy difference between rotational levels with different $J$ \cite{schmidt06,leibfried12,ding12}. Rather than relying on blackbody radiation to randomly populate a level with the desired $J$, this should allow us to coherently transfer the molecule from any pure state to the desired state on time scales much faster than the blackbody rates~\cite{schmidt06,leibfried12,ding12}. We are optimistic that precise spectroscopic experiments on a multitude of molecular ions that may be relevant in astrophysics, stringent tests of fundamental theories as well as quantum information processing will be enabled by these techniques.

\section{Acknowledgements}

We thank K. C. Cossel, Y. Wan, and D. J. Wineland for helpful comments on the manuscript. This work was supported by the U.S. Army Research Office and the NIST quantum information program.  C. Kurz acknowledges support from the Alexander von Humboldt foundation. P. N. Plessow acknowledges support by the state of Baden-W\"{u}rttemberg through bwHPC. Contribution of National Institute of Standards and Technology, not subject to U.S. copyright.

\section{Author contributions}
C.W.C. and D.L. conceived and designed the experiments. C.W.C.~and C.K. developed components of the experimental apparatus and collected and analyzed data. C.W.C. and D.L. wrote the manuscript. D.B.H. and D.R.L. contributed to the development of experimental methods and pulse sequences. P.N.P. computed the molecular constants and level structure. All authors provided important suggestions for the experiments, discussed the results, and contributed to the editing of the manuscript.

\section{Methods}

\subsection{Additional references}
Due to the limit imposed on the number of references for the main text, we add references here for a more comprehensive survey of the field. The breakthroughs enabled by  laser cooling and trapping of atoms and atomic ions in realizing excotic phases of matter are summarized in~\cite{BEC1,BEC2,fermi1,fermi2,ioncrystal}. Opportunities offered by precision experiments with cold trapped molecules are explored~\cite{carr09,demille15} for precision metrology~\cite{koelemeij07,schiller05,flambaum07} and quantum information science~\cite{pupillo09,demille02}. In addition to coherent optical control~\cite{ospelkaus10}, recently, coherent microwave control of ultracold molecules was demonstrated~\cite{will16}. Some of the schemes proposed to reduce number of states occupied by molecular ions can be found in~\cite{schmidt06,hudson08,lazarou10,murpetit12,shi13,vogelius06}, with notable similarity between the scheme proposed in~\cite{vogelius06} and that demonstrated in our work. Forming \CaH~via laser-induced reaction and sympathetic cooling of \CaH~with \Ca~ions was studied in~\cite{kimura11} and ground state cooling of a \CaH--\Ca pair was reported in~\cite{rungango15}.

\subsection{Generality of our approach and special applications}

Sympathetic cooling and quantum logic readout require that the motion of the molecular ion and the logic ion are sufficiently coupled \cite{leibfried12}. The amplitude of the heavier ion decreases and the frequency difference of the two normal modes increases as the mass ratio of the constituents becomes more unequal. An additional factor is that the radial pseudo-potential of linear rf-traps scales with the inverse mass, therefore the disparity in the radial frequencies also increases when masses become more different. Quantum logic experiments have been done at mass ratios of up to three \cite{schmidt05} and this ratio could probably be pushed higher. Examples of well characterized logic ions are $^9$Be$^+$, $^{25}$Mg$^+$, $^{40}$Ca$^+$, $^{88}$Sr$^+$, $^{138}$Ba$^+$ and $^{171}$Yb$^+$. Use of these species should allow a range of diatomic and polyatomic molecules with masses from 3 amu to 513 amu, sufficiently high to co-trap ions of small organic compounds. Despite the fact that their mass ratio is larger than three, we are optimistic that the lightest molecular ion, H$_2^+$ can be made accessible with $^9$Be$^+$ as the logic ion.\\
\\
The single continuous wave laser at 1051 nm used in this work, or similar continuous wave sources can address Raman transitions between states with energy differences ranging from a few 100 Hz to tens of GHz, where the two Raman beams can be separated in frequency with suitable acousto-optic or electro-optic devices, as demonstrated in our work. In this way one can address energy levels split by a few to several tens of kHz that arise from the coupling of nuclear magnetic moments to those induced by molecular rotation and due to coupling of such magnetic moments to weak external magnetic fields similar to the states used in our manuscript. Energy splittings due to coupling of the electron spins with the rotation, hyperfine and Zeeman structure that typically range between a few MHz and 10 GHz can also be bridged in this manner.\\
\\
Stimulated Raman transitions with reasonable Rabi-frequencies are feasible in symmetric molecules, despite their lack of a permanent dipole moment. This is also true for H$_2^+$ that has no stable excited electronic state. With sufficiently high detuning, the population in excited states can be kept so small, that the probability of dissociation becomes negligible. Nevertheless, the electric fields due to the Raman beams deform the electronic ground state in such a way that it has a polarization component at energy differences that arise within the electronic ground state manifold due to vibration, rotation, hyperfine structure and so forth. Due to its unbound excited electronic state, the sum over excited states that we evaluate below cannot be used for computing the Raman-Rabi frequency in H$_2^+$. However, the Rabi frequency can be computed with other methods \cite{karr14}. For light around 800 nm with 750 mW per Raman-beam and focused to a 20 $\mu$m$^2$ waist, the Rabi frequency can be estimated to be $2 \pi \times$ 29 kHz for a carrier transition \cite{hilico17}.\\
\\
The search for an electron electric dipole moment (EDM) could be realized by combining our methods with those described in \cite{loh13}. This experiment uses a linear trap with extra electrodes to superimpose a rotating radial electric field onto the trapping fields. If the radial electric field rotation frequency is much faster than the radial secular frequency of the ions,  it leads to additional circular micro-motion at the frequency of the rotating field. The ions micro-motion excursions are essentially 180 degrees out of phase with the rotating electric field, so they experience a field of constant magnitude, but with a zero average over one rotation period. Due to couplings between the electron spin and the molecular rotation, the lowest energy splittings (due to $\Lambda$ doubling~\cite{loh13}) that dictate adiabaticity are around 10~MHz. In the experiments described in~\cite{loh13}, the field has magnitude 11.6 V/cm and rotates at 253 kHz. In a linear trap with 2~mm distances between opposing electrodes, such a rotating field could be achieved with less than 3 V of amplitude, considerably below amplitudes commonly used for producing the radial confinement in linear traps of such dimensions (tens to hundreds of V). However, the extent of the circular motion of the ions will be sizable (order 200 $\mu$m) and the secular frequencies will be limited to tens of kHz. Quantum logic spectroscopy at such low secular frequencies has not been demonstrated so far. The rotating field does not need to be turned on during the quantum logic interrogation. If all technical obstacles can be overcome, the question still remains whether experiments of a single or few molecular ions will offer an advantage over the larger ensembles that are currently probed in the experiments described  in \cite{loh13}. For that, the advantage in ion number needs to be overcompensated by an increased coherence time and precision. It seems conceivable that this could be the case, because the coherence time demonstrated in \cite{loh13} was likely limited by  uncontrolled ion-ion interactions. Moreover, the quantum logic setup may support having different molecular ions side by side with the logic ion and non-destructively probing them in the same environment. This might be advantageous to further suppress systematic uncertainties.\\
\\
Another field of current interest is to use molecular transitions to increase the sensitivity of searches for the variation of fundamental constants. Vibrational transitions are most sensitive to variations in the ratio of electron to proton mass $m_e/m_p$ and probing them will require dedicated lasers. Nevertheless, the methods demonstrated here can accomplish efficient state preparation and non-destructive state read-out with near unity quantum efficiency and thus overcome major hurdles on the way to such experiments. The species used in our experiments, CaH$^+$, has been identified as a candidate for testing the time dependence of $m_e/m_p$ as well as other earth-alkali halides  (SrH$^+$, YbH$^+$) \cite{kajita14a} and NH$^+$ \cite{beloy11} that have a similar level structure. Manipulation of H$_2^+$ and HD$^+$ \cite{schiller05} is discussed in more detail above. The symmetric molecules N$_2^+$ \cite{kajita14b} and O$_2^+$ \cite{hanneke16} have also been suggested as viable candidates with little coupling to the environment, due to the lack of a permanent dipole moment. Both are a good match for Ca$^+$ as the logic ion and have magnetic dipole moments that allow to project and measure them with a continuous  wave source and  suitable acousto-optic or electro-optic devices as described above. The species discussed in \cite{pasteka15} (HBr$^+$, HI$^+$, Br$_2^+$, I$_2^+$, IBr$^+$, ICl$^+$, IF$^+$) all have a $\Pi$ electronic ground state that leads to a magnetic moment due to the orbital angular momentum of the electrons that will provide accessible transitions, with more level splittings provided by additional magnetic moments. However, these ions are heavier, so Ba$^+$ or Yb$^+$ might be a more suitable logic ion.

\subsection{Numerical Calculation of $\text{CaH}^+$ Properties}
Calculating the properties of CaH$^+$ requires an accurate potential energy curve to determine the equilibrium structure and rovibrational wave functions within the Born-Oppenheimer approximation.
In a second step, properties (spin-rotation constant and g-factor) are computed as a function of the Ca-H internuclear distance $r$.
The g-factor and spin-rotation constant for a given rovibrational state $|Jv \rangle$ are computed as the expectation value with respect to the rovibrational wavefunction, for example $g_{Jv}=\langle\Psi_{Jv}(r) |g(r) |\Psi_{Jv}(r)\rangle$.
\subsubsection{Geometry and Vibrational Wave Function of CaH$^+$}
\begin{figure}[H]
\includegraphics[angle=-0,width=0.7\textwidth]{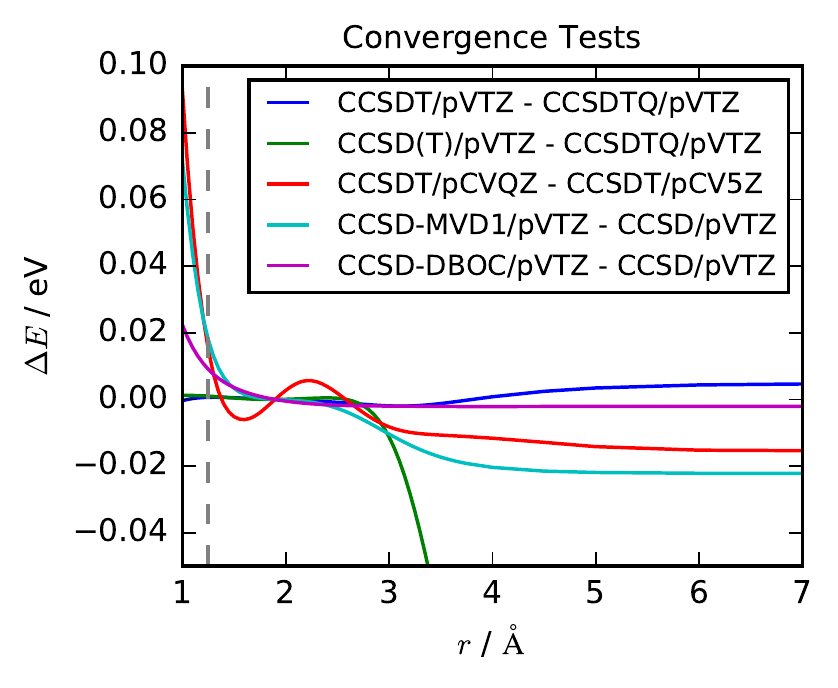}%
\centering
\caption{Difference in energies obtained with different methods.
All curves have been shifted to be zero at 1.896 $\mathrm{\AA}$.
The grey dashed line is drawn at 1.25 $\mathrm{\AA}$ which is the lower limit of a significant probability density at a high vibrational quantum number ($v$=10).
Basis sets are correlation-consistent basis sets (the cc has been omitted for clarity).
MVD1 and DBOC donate mass-velocity-1-electron-Darwin and diagonal Born-Oppenheimer corrections. At $r$ $>$ 3.0 {\AA}, which is outside of the range that is relevant for the electronic and vibrational ground state,  CCSD(T) starts to deviate significantly from CCSDT and deviations lie outside of the energy scale.
\label{method_conv}}
\end{figure}

Calculations have been carried out with the program packages CFOUR~\cite{cfour} and MRCC~\cite{mrcc} using coupled cluster (CC) methods.
Different levels of coupled cluster theory are abbreviated according to the level of excitations in the exponential, e.g. CCSDTQ stands for coupled-cluster with single, double, triple and quadruple excitations.
In CCSD(T), the contribution of triple excitations is determined from perturbation theory~\cite{pople1987quadratic}.
All calculations employ atom-centered Gaussian basis sets, the correlation-consistent polarized basis sets (cc-p)~\cite{RN272,RN273}, with valence-only (cc-pV) and core-valence (cc-pCV) correlation for Ca.
The full basis set is specified by the number X of independent radial basis functions per correlated occupied orbital (XZ), for example cc-pCV5Z.
All calculations are based on a closed-shell restricted Hartree-Fock reference.
The frozen-core approximation has been used in coupled cluster calculations, with the 5 lowest (doubly occupied) orbitals in the frozen core.
In the following we are not interested in total energies but only in the shape of the potential energy curve.
As can be seen in Figure \ref{method_conv}, the energy is well converged with respect to the cluster operator at the CCSDT level of theory.
This follows from comparison with CCSDTQ-calculations with deviations below 0.01 eV.
In the range of the potential curve that is relevant for the vibrational ground state (approximately 1.4 {\AA} $<$ $r$ $<$ 2.4 {\AA}, see Figures \ref{c-distance} and \ref{g-distance}), CCSD(T) agrees furthermore quantitatively with CCSDT.
At $r$ $>$ 3.0 {\AA}, CCSD(T) starts to deviate significantly from CCSDT since the calculation of the triples amplitudes from perturbation theory is no longer a good approximation to the solution of the coupled cluster equations.
We therefore use CCSD(T) since it is accurate enough and most efficient.
Increasing the basis set from cc-pCVQZ to cc-pCV5Z results in energy changes below 0.02 eV.
The contribution of the diagonal Born-Oppenheimer correction (DBOC) and scalar-relativistic mass-velocity-1-electron-Darwin (MVD1) corrections are also on the order of 0.02 eV.

Vibrational wavefunctions are obtained by numerical solution of the Schr{\"o}dinger equation using the full potential $V_0(r)$ (energy of the electronic wavefunction as a function of Ca-H distance, using CCSD(T)) plus centrifugal potential:
\begin{align}
V(r) = V_0(r) + \frac{\hbar^2 J(J+1)}{2\mu r^2},
\end{align}
where $\mu$ is the reduced mass and $J$ is the rotational quantum number.
\subsubsection{Hyperfine-coupling and g-factor from numerical electronic structure theory}
Spin-rotation constants and g-factors have been computed from coupled-cluster response theory using London atomic orbitals and rotational London atomic orbitals~\cite{RN254,RN266,RN269,RN264}.
As has been found before, the calculations are not very sensitive to basis set size or core-correlation~\cite{RN254}.
At the minimum of the potential energy curve, computed values for $g$ and $c_{IJ}$ deviate by about 0.1\% comparing CCSDT and CCSD(T) and still by less than 0.5\% comparing CCSDT and CCSD, demonstrating well-converged results with respect to the cluster operator (see Table \ref{tab_theory1}).
Results obtained with the 3$\zeta$ basis set cc-pVTZ differ significantly from those obtained with larger basis sets, (13\% for $g$).
The difference between the calculation with the largest and second-largest basis set (CCSD(T)/cc-pCV5Z and CCSD(T)/cc-pCVQZ) is 1.7\% for $g$ and 0.6\% for $c_{IJ}$.
In the literature, computed values based on a comparable level of theory deviate usually by less than 5~\% from experimental values, and often the deviation is less than 1~\%~\cite{RN263,RN259,RN254,RN255}.
Based on these results, we will use CCSD(T)/cc-pCV5Z to compute all properties and estimate the error of computed values of $c_{IJ}$ and $g$ to be $\pm$5\%. The minimum of the potential energy curve at the CCSD(T)/cc-pCV5Z level of theory is $r_0$ = 1.896 \AA, in agreement with~\cite{abe2012}.
\begin{table}
\caption{Numerical values of spin-rotation constant $c_{IJ}$ and g-factor $g$ at $r_0$ = 1.896 \AA, which is the minimum of the potential energy curve at the CCSD(T)/cc-pCV5Z level of theory.}
\begin{tabular}{llrr}
\\
\hline
\hline
method&basis set& \multicolumn{1}{c}{$g$} &  \multicolumn{1}{c}{$c_{IJ}$/kHz}
\\
\hline
  CCSD(T) & cc-pCVQZ & -1.38 & 8.47
\\CCSD(T) & cc-pCV5Z & -1.36 & 8.52
\\CCSD(T) & cc-pV5Z &  -1.37 & 8.50
\\CCSD(T) & cc-pVTZ &  -1.18 & 8.67
\\CCSDT   & cc-pVTZ &  -1.19 & 8.67
\\CCSD    & cc-pVTZ &  -1.18 & 8.71
\\
\hline
\hline
\end{tabular}
\label{tab_theory1}
\end{table}

Vibrationally averaged values are obtained using the vibrational wave-function as well as a polynomial interpolation of the property.
Relativistic corrections to the spin-rotation constant from Thomas precession~\cite{RN258} have been included but are always small, being on the order of 10 Hz.
The results are presented in Table \ref{tab_theory}.
The vibrational corrections are smaller for the g-factor, because $g(r)$ is approximately linear in the relevant range of $r$.
Since the vibrational ground state wavefunction is to a good approximation as symmetric around the equilibrium distance $r_0$ as the harmonic approximation, the vibrational correction to any property that depends linearly on $r$ nearly vanishes.
\begin{table}
\caption{Numerical values of spin-rotation constant $c_{IJ}$ and g-factor $g$ in the vibrational ground state $v=0$ and in the 15 lowest rotational states.}
\begin{tabular}{lrr}
\\
\hline
\hline
& \multicolumn{1}{c}{$g$} & \multicolumn{1}{c}{$c_{IJ}$/kHz}
\\
\hline
   $J=0 $        & -1.35  & 8.27
\\ $J=1 $        & -1.35  & 8.26
\\ $J=2 $        & -1.35  & 8.26
\\ $J=3 $        & -1.34  & 8.26
\\ $J=4 $        & -1.34  & 8.26
\\ $J=5 $        & -1.34  & 8.25
\\ $J=6 $        & -1.34  & 8.25
\\ $J=7 $        & -1.34  & 8.24
\\ $J=8 $        & -1.33  & 8.24
\\ $J=9 $        & -1.33  & 8.23
\\ $J=10$        & -1.33  & 8.22
\\ $J=11$        & -1.32  & 8.21
\\ $J=12$        & -1.32  & 8.20
\\ $J=13$        & -1.31  & 8.19
\\ $J=14$        & -1.31  & 8.18
\\
\hline
\hline
\end{tabular}
\label{tab_theory}
\end{table}

\begin{figure}[H]
\centering
\includegraphics[angle=-0,scale=1.300]{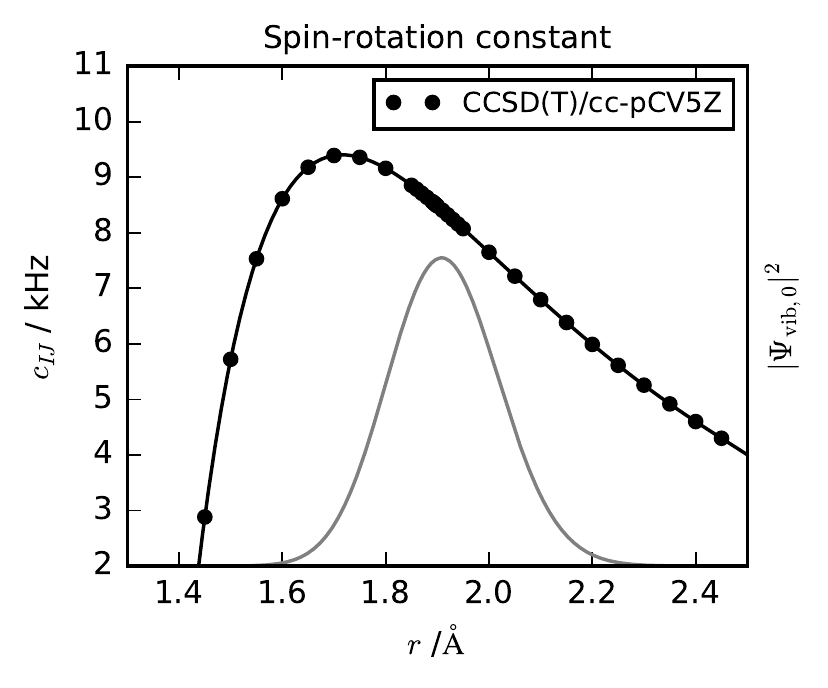}%
\caption{Spin-rotation factor, fit with an 8th order polynomial.
The squared vibrational wavefunction is shown in grey for $J=0$.\label{c-distance}}
\end{figure}

\begin{figure}[H]
\centering
\includegraphics[angle=-0,scale=1.300]{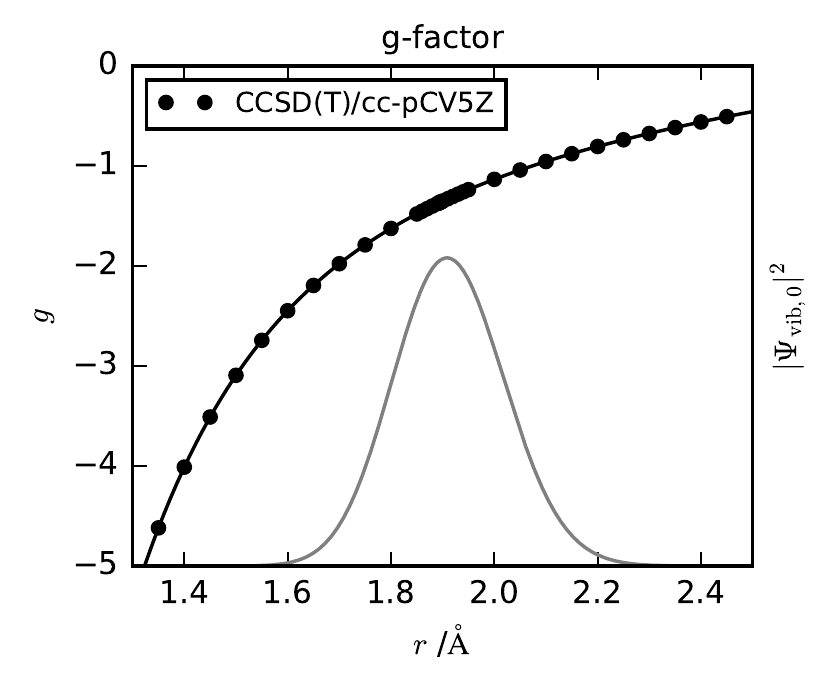}%
\caption{g-factor, fit with an 8th-order polynomial.
The squared vibrational wavefunction is shown in grey for $J=0$.\label{g-distance}}
\end{figure}

The behavior of the spin-rotation constant and g-factor as a function of the bond distance is shown in Figures \ref{c-distance} and \ref{g-distance}.

\subsection{Magnetic sublevels of $\text{CaH}^+$ \label{SecMagSub}}

The energies of the rotational levels of the molecule referenced to the electronic and vibrational ground state are approximately proportional to the eigenvalues of the square of the rotational angular momentum operator, $\hat{\mathbf{J}}^2$, namely $h R~ J(J+1)$. For a given rotational quantum number $J$ and with a magnetic field $\mathbf{B}$, the energy of a state is further determined by the mutual (hyperfine) coupling between the orbital magnetic moment, produced by the charges rotating with the molecule, and the magnetic moment of the hydrogen nucleus ($\propto \hat{\mathbf{J}}\cdot \hat{\mathbf{I}}$), as well as the coupling of the individual magnetic moments to the external magnetic field ($\propto \hat{\mathbf{J}}\cdot \mathbf{B}$ and $\propto \hat{\mathbf{I}}\cdot \mathbf{B}$) which split the levels by a few kilohertz. With the spin-rotation constant $c_{IJ}$ and g-factor $g$, the structure of \CaH~magnetic sublevels in each $J$-manifold can be calculated by diagonalizing the Hamiltonian
\begin{equation}
\hat{H }= -g \frac{\mu_{N}}{\hbar}\hat{\mathbf{J}}\cdot \mathbf{B}-g_I \frac{\mu_{N}}{\hbar}\hat{\mathbf{I}}\cdot \mathbf{B}-\frac{2\pi c_{IJ}}{\hbar} \hat{\mathbf{I}}\cdot \hat{\mathbf{J}}\text{,}
\end{equation}
where $g_I$ is the proton g-factor and $\mu_N$ is the nuclear magneton. In our experiments, the external magnetic field defines the quantization axis, $\mathbf{B}=B \mathbf{e}_{\tilde{z}}$, where $\mathbf{e}_{\tilde{z}}$ is a unit vector along the magnetic field direction (the $\tilde{z}$-axis does not coincide with the $z$-axis along which the ion crystal aligns, see Fig. 1 of the main text) and $B \simeq$ 0.36~mT. The molecule is neither in the Zeeman regime, where the mutual coupling of the spins is stronger than their coupling to the external field, nor the Paschen-Back regime where the spins tend to align with the magnetic field. Defining the eigenvalues of $\hat{J}_{\tilde{z}}+\hat{I}_{\tilde{z}}$ as $\hbar m=\hbar(m_J+ m_I)$ and using the operator indentity $\hat{J}_{\tilde{x}} \hat{I}_{\tilde{x}}+\hat{J}_{\tilde{y}} \hat{I}_{\tilde{y}} =1/2(\hat{J}_+ \hat{I}_-+ \hat{J}_+ \hat{I}_-)$ where $\hat{J}_\pm |J,m_J\rangle=\hbar\sqrt{(J\mp m_J)(J\pm m_J+1)} |J, m_J\pm 1\rangle$ act as ladder operators, as long as $|m|\leq J$ (and likewise $\hat{I}_\pm$), we can rewrite
\begin{equation}
\hat{H} = -(g \hat{J}_{\tilde{z}}  +g_I  \hat{I}_{\tilde{z}})\frac{\mu_{N}}{\hbar} B-\frac{2\pi c_{IJ}}{\hbar} [\hat{J}_{\tilde{z}} \hat{I}_{\tilde{z}} +1/2(\hat{J}_+ \hat{I}_- + \hat{J}_- \hat{I}_+)] ,
\end{equation}
which shows that irrespective of the value of $B$, $m$ is a good quantum number, since all terms in $\hat{H}$ preserve $m$. For \CaH, the nuclear spin of the proton is $I=1/2$ and the Hamiltonian is block-diagonal in the basis of product states $|J, m_J\rangle|I, m_I\rangle$. For the extreme cases $m=\pm(J+1/2)$, the eigenstates of $\hat{H}$ are $|J,\pm (J+1/2),\pm\rangle=|J, \pm J \rangle |1/2, \pm1/2\rangle$, fully aligned product-states with energies $E_{\pm (J+1/2),\pm}=\mp (g J +g_I/2)\mu_{N} B \mp h c_{IJ} J/2 $. In all other cases, the blocks have dimensions $2\times2$ and the form
\begin{equation}
\hat{H}_{J,m} =
\left(
\begin{array}{ccc}
 -\mu_N B [g (m - \frac{1}{2}) + \frac{g_I}{2}] - h \frac{c_{IJ}}{2} (m - \frac{1}{2})&  - h \frac{c_{IJ}}{2} \sqrt{(J+\frac{1}{2})^2-m^2} \\
-h \frac{c_{IJ}}{2} \sqrt{(J+\frac{1}{2})^2-m^2}  &  -\mu_N B [g (m + \frac{1}{2}) - \frac{g_I}{2}] + h \frac{c_{IJ}}{2} (m + \frac{1}{2})
\end{array}
\right).
\end{equation}
When diagonalizing such a block in analogy to the solution of the Breit-Rabi equation we get the eigenvectors
\begin{eqnarray}\label{EqEigVec}
|J,m, +\rangle &=&\sqrt{\frac{X-Y}{2 X}}~|J, m-1/2 \rangle |1/2, 1/2\rangle +\sqrt{\frac{X+Y}{2 X}} |J, m+1/2 \rangle ~|1/2, -1/2\rangle\text{  and}\nonumber \\
|J,m, -\rangle &=&-\sqrt{\frac{X+Y}{2 X}} ~|J, m-1/2 \rangle |1/2, 1/2\rangle+\sqrt{\frac{X-Y}{2 X}} ~|J, m+1/2 \rangle |1/2, -1/2\rangle,
\end{eqnarray}
with
\begin{eqnarray}
X&=&\frac{1}{2}\sqrt{h^2 c_{IJ}^2 [(J+\frac{1}{2})^2 - m^2] + [h c_{IJ} m - \mu_N B (g - g_I)]^2}\text{ and} \nonumber\\
Y&=&-\mu_N B( \frac{g}{2} - \frac{g_I}{p}) - m h \frac{c_{IJ}}{2}.
\end{eqnarray}
The corresponding eigenvalues are
\begin{equation}
E_{m,\pm}=h \frac{c_{IJ}}{4} - \mu_N B g~ m  \mp  X.
\end{equation}
The contribution from $X$ can approximately cancel the other factors in $E_{m,+}$, leading to small and similar transition frequencies between states with different $m$ in that manifold, while the $m$-dependent effects add up in $E_{m,-}$ (shown for $J=1,2$ in Fig. 2 of the main text). We take advantage of small frequency differences in both manifolds to pump several substates simultaneously, while using the most distinguishable transitions $|J,-J+1/2,-\rangle \leftrightarrow |J,-J-1/2,-\rangle$ for state determination and spectroscopy (see also the sections on pumping and spectroscopy below). To simplify expressions containing the eigenstates in the main text, we use the abbreviation $|\mathcal{J}\rangle$ as a shorthand for states with the set of quantum numbers $|J,m,\xi \rangle$ with $\xi \in \{+,-\}$.
\subsection{Stimulated Raman transitions\label{SRS}}
\begin{figure}[H]
\includegraphics[angle=-0,scale=0.400]{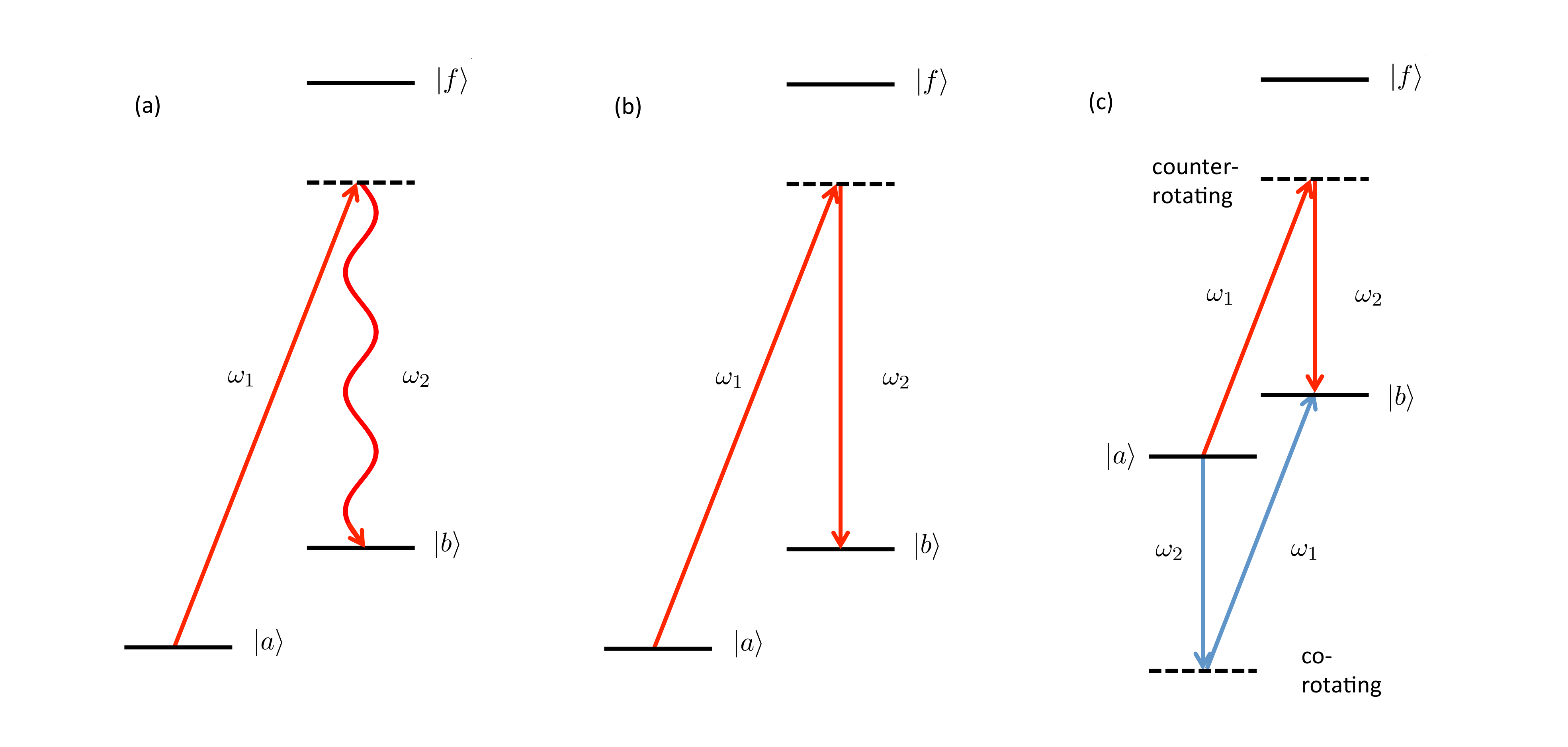}%
\caption{\label{FigRamTra} Raman transitions: (a) Single field Raman transition, photon at $\omega_2$ is spontaneously emitted. (b) Stimulated Raman transition, driven by two fields, photon at $\omega_2$ is stimulated by the second driving field. (c) Order of absorption and emission and detuning of the coherence in the counter-rotating (red) and co-rotating (blue) cases.}
\end{figure}

Raman scattering on molecules usually involves a single light field at $\omega_1$ that is far off-resonant from all intermediate excited states $|f\rangle$. Then the second, Stokes or Anti-Stokes photon at $\omega_2$ is spontaneously emitted by the molecule as shown in Fig. \ref{FigRamTra} (a).  In our experiments, we excite transitions from $|a\rangle$ to $|b\rangle$ where the second photon at $\omega_2$ is stimulated by a second driving field at that frequency, as shown schematically in Fig. \ref{FigRamTra} (b). The second field greatly enhances the Raman scattering rate and leads to coherent transitions between $|a\rangle$ and $|b\rangle$. The two driving light fields are characterized by ($n=\{1,2\}$) frequency $\omega_n$, field amplitude $|\mathbf{E}_n|$ and polarization $\hat{\mathbf{q}}^{(n)}=q^{(n)}_{-1}\mathbf{\sigma}^-+q^{(n)}_0 \mathbf{\pi}+  q^{(n)}_1\mathbf{\sigma}^+$,  $|\hat{\mathbf{q}}^{(n)}|^2=1$, where $\pi$ is oriented along the quantizing magnetic field and $\mathbf{\sigma}^+$ ($\mathbf{\sigma}^-$) is circular polarization rotating clockwise (counter-clockwise) around that direction when viewed along the wavevector. The fields are far off-resonant from the smallest frequency difference between states of the molecule with the electrons in the ground state and those with electrons in excited states.

While each light field is very far detuned from resonance, we assume that the frequency difference $\Delta \omega=(\omega_2-\omega_1)$ is close to the frequency difference $\omega_0=1/\hbar(E_b-E_a)$ of the initial molecular state $|a\rangle$ and a final state $|b\rangle$. Then, the Raman-Rabi-frequency $\Omega_{ab}$ for a stimulated Raman transition can be approximated in perturbation theory as
\begin{equation}\label{EquRamRab}
\Omega_{ab}=\frac{1}{4 \hbar^2}\sum_{f} \frac{\langle b|e~ \mathbf{r}\cdot \mathbf{E}_2|f\rangle\langle f | e~ \mathbf{r}\cdot\mathbf{E}_1|a\rangle}{\omega_{af}-\omega_1}+
\frac{\langle b|e~ \mathbf{r}\cdot \mathbf{E}_1|f\rangle\langle f | e~ \mathbf{r}\cdot\mathbf{E}_2|a\rangle}{\omega_{af}+\omega_2},
\end{equation}
where $e$ is the charge of the electron, $\mathbf{r}$ its position and $\omega_{af}=1/\hbar(E_f-E_a)$ is the frequency difference between the initial state $|a\rangle$ and an intermediate state $|f\rangle$. The term in the sum with denominator $\omega_{af}-\omega_1$ is due to terms rotating at the difference frequency between the first light field and the energy difference of initial and intermediate state. In most textbook discussions, the difference is assumed to be much smaller than the sum $\omega_{af}+\omega_2$ that appears in the denominator of the other term. In these cases, terms in the perturbation expansion co-rotating at this sum are neglected with respect to the counter-rotating terms at the difference frequency in what is called the rotating wave approximation. However, in our experiments, the light fields are detuned so far that the ratio of difference to sum is approximately 1:2.3 for laser light near 1051 nm and estimating the energy difference of the ground and first excited electronic levels according to~\cite{abe2012}. Therefore, we cannot necessarily neglect the co-rotating terms. The difference in the processes is illustrated in Fig. \ref{FigRamTra} (c): For the counter-rotating terms (red arrows), the photon at $\omega_1$ is absorbed to set up a coherence in the molecule that rotates at $\omega_{af}-\omega_1$, then a photon at $\omega_2$ is stimulated. In the co-rotating term (blue arrows), the photon at $\omega_2$ is stimulated first, setting up a coherence at the larger detuning $\omega_{af}+\omega_2$, then $\omega_1$ is absorbed, leading to the same resonance condition and selection rules as in the counter-rotating term.\\
\\
In our experiments, initial and final states are in the electronic ground state, $|1^1\Sigma\rangle$ in the notation of~\cite{abe2012}, vibrational ground state, $v=0$, and a particular manifold of the rotation characterized by quantum number $J$. To determine the Raman-Rabi-rate, we need to evaluate dipole matrix elements of the form
\begin{equation}\label{EquDipMat}
\langle f | e~ \mathbf{r}\cdot\mathbf{E}|s\rangle=e |\mathbf{E}| \langle  \Psi_{f,v}(r)|r|1^1\Sigma, v=0 \rangle \langle\Phi_f(\theta, \phi) | \hat{\mathbf{r}}\cdot\hat{\mathbf{q}}|\Phi_s(\theta, \phi)\rangle,
\end{equation}
where $s\in\{a,b\}$. On the right-hand side we have factored the molecular wave functions into two parts. The radial part of the excited state electronic wave-function $\Psi_{f,v}(r)$ only depends on the inter-nuclear distance $r=|\mathbf{r}|$ and vibrational quantum number $v$. The angular part of the wavefunction $\Phi_s(\theta, \phi)$ only depends on the two angle coordinates $\theta$ and $\phi$. Such a separation of variables can be justified, for example, within the Born-Oppenheimer approximation. Here, we only consider the lowest excited electronic state $|2^1\Sigma\rangle$ and neglect the differences in $\omega_{af}$ due to the rotational and vibrational states, $\omega_{af}\simeq 1/\hbar(E_{2^1\Sigma}-E_{1^1\Sigma})\equiv\bar{\omega}$. Under these assumptions we can rewrite Eq.(\ref{EquRamRab}) as
\begin{eqnarray}\label{EquRamFac}
\Omega_{ab}&=& \mathcal{E} (S_-+S_+) \nonumber \\
\mathcal{E} &=& \frac{e^2 |\mathbf{E}_1| |\mathbf{E}_2|}{4 \hbar^2 (\bar{\omega}-\omega_1)}\sum_{v'} |\langle \Psi_{2^1\Sigma,v'}|r|\Psi_{1^1\Sigma,0}\rangle|^2,\nonumber \\
S_-&=&\sum_f \langle \Phi_b|\hat{\mathbf{r}}\cdot \hat{\mathbf{q}}^{(2)}|\Phi_f\rangle\langle \Phi_f | \hat{\mathbf{r}}\cdot\hat{\mathbf{q}}^{(1)}|\Phi_a\rangle\nonumber\\
S_+&=&\frac{\bar{\omega}-\omega_1}{\bar{\omega}+\omega_1} \sum_f \langle \Phi_b|\hat{\mathbf{r}}\cdot \hat{\mathbf{q}}^{(1)}|\Phi_f\rangle\langle \Phi_f | \hat{\mathbf{r}}\cdot\hat{\mathbf{q}}^{(2)}|\Phi_a\rangle ,
\end{eqnarray}
where we have also neglected the small frequency differences between the driving fields $\omega_1\simeq\omega_2$ in the denominators. Under these approximations, $\mathcal{E}$ is a prefactor that is the same for all transitions considered here and is proportional to the sum over the squares of all Franck-Condon factors of vibrational states bound in the well of the first excited electronic state.\\
\\
The position operator $\mathbf{r}$ couples eigenstates of the rotational angular momentum $|J,m_J\rangle$ and has no effect on the eigenstates of the proton nuclear spin $|I, m_I\rangle$. Therefore, it is convenient to use the product basis $|J, m_J\rangle| I, m_I\rangle$ with the nuclear spin $I$ fixed. In $^{40}$CaH$^+$, the nuclear spin $I=1/2$ is that of the proton. The sum over the intermediate states $\Phi_f$ in $S_-$ is
\begin{equation}\label{EquSumAng}
S_-=\sum_{J,m_J,m_I} \langle \Phi_b| \hat{\mathbf{r}}\cdot \hat{\mathbf{q}}^{(2)}|J,m_J\rangle|I,m_I\rangle\langle J,m_J|\langle I,m_I| \hat{\mathbf{r}}\cdot\hat{\mathbf{q}}^{(1)}|\Phi_a\rangle
\end{equation}
The sum of angular momenta in the z-direction $m_s=m_J+m_I$ remains a good quantum number for all values of the magnetic field $\mathbf{B}=B \hat{\mathbf{e}}_z$. This implies that we can write the eigenstates as
\begin{equation}\label{EquEigSta}
|\Phi_s\rangle = \sum_{m_I} c^{(s)}_{m_I} |J_s,m_{s}-m_I,1/2,m_I\rangle,~ s\in\{a,b\},
\end{equation}
where $c^{(s)}_{m_I}$ are the coefficients of the eigenstates derived in Eq.(\ref{EqEigVec}). Inserting Eq.(\ref{EquEigSta}) into Eq.(\ref{EquSumAng}) yields
\begin{equation}\label{EquSumAng2}
S_-=\sum_{J,m_J,m_I} c^{(b)}_{m_I} c^{(a)}_{m_I}\langle J_b, m_{b}-m_I| \hat{\mathbf{r}}\cdot \hat{\mathbf{q}}^{(2)}|J,m_J\rangle \langle J,m_J| \hat{\mathbf{r}}\cdot\hat{\mathbf{q}}^{(1)}|J_a, m_{a}-m_I\rangle.
\end{equation}
We can express $\hat{\mathbf{r}}$ as a vector in the spherical basis to write out the scalar product and use the Wigner-Eckart theorem  to determine the matrix element for photon absorption from a field with polarization $\hat{\mathbf{q}}$:
\begin{equation}\label{EquMatAbs}
\langle J, m|\hat{\mathbf{r}}\cdot\hat{\mathbf{q}}|J', m'\rangle=\sqrt{{\rm Max}(J,J')} ~(J-J')\sum_{k=-1}^1 \hat{q}_k (-1)^{k+J-m}
\left(
\begin{array}{ccc}
  J &   1 &  J' \\
  -m&  k & m'
\end{array}
\right),
\end{equation}
where the 2x3 array in brackets at the end is a Wigner 3J-symbol. The right hand side implies the usual dipole selection rules which reduce the number of excited states that need to be considered in the Raman Rabi-frequency as well as the final states that can have a non-zero coupling to $|\Phi_a\rangle$. In particular, $\Delta J = J_b-J_a =0,\pm2$ and $\Delta m=m_b-m_a=0,\pm1,\pm2$. Restricting the sums with the selection rules and setting $m_{J_a}=m_{a}-m_I$, we get
\begin{eqnarray}\label{EquSumAng3}
S_-&=&\sum_{m_I=-1/2}^{1/2} ~~\sum_{k_1,k_2=-1}^1  c^{(b)}_{m_I} c^{(a)}_{m_I}  \left[ \langle J_a-1,m_{J_a}+k_1| \hat{r}_{k_2} q^{(2)}_{k_2}|J_b, m_{J_a}+k_1-k_2\rangle \times \right.\nonumber \\
& & \langle J_a-1,m_{J_a}+k_1| \hat{r}_{k_1} q^{(1)}_{k_1}|J_a, m_{J_a}\rangle +\nonumber\\& &\langle J_a+1,m_{J_a}+k_1| \hat{r}_{k_2} q^{(2)}_{k_2}|J_b, m_{J_a}+k_1-k_2\rangle \times \nonumber\\
& &\left. \langle J_a+1,m_{J_a}+k_1| \hat{r}_{k_1} q^{(1)}_{k_1}|J_a, m_{J_a}\rangle \right],
\end{eqnarray}
with the matrix elements all written in the form where the photon is absorbed to be compatible with Eq.(\ref{EquMatAbs}). The expressions for $S_+$ are calculated in the same way, but the available intermediate states change since the photon from field $\mathbf{E}_2$ and its angular momentum is absorbed first in those terms.\\
\\
AC-Stark shifts arise due to Raman transitions with $|a\rangle=|b\rangle$ and amount to an energy shift of state $|a\rangle$ by $\Delta E_{AC}=\hbar \Omega_{aa}$. If the two driving fields have different frequencies and/or polarizations, both photons contributing to a Raman transition need to come from the same field to conserve energy and angular momentum. For two fields with polarizations $\hat{\mathbf{q}}^{(1)}, \hat{\mathbf{q}}^{(2)}$ and setting $J_a = J$, $m_a=m$ and $c_{m_I}^{(a)}=c_{m_I}$ we can simplify Eq.(\ref{EquRamFac}) to
\begin{eqnarray}\label{EquSumAC}
\Delta E_{AC}=({\mathcal E}_1+{\mathcal E}_2)S_{-} \frac{2 \bar{\omega}}{\bar{\omega}-\omega_1} =
 \frac{2 \bar{\omega}}{\bar{\omega}-\omega_1} \sum_{n=1}^2 {\mathcal E}_n \sum_{m_I=-1/2}^{1/2}  |c_{m_I}|^2 \times& & \nonumber \\
\left[ \frac{(m+m_I)^2(3 |q_{0}^{(n)}|^2-1)}{3-4 J(J+1)}+\frac{1-J(J+1)(|q_{0}^{(n)}|^2+1)}{3-4 J(J+1)} \right].
\end{eqnarray}
The AC-Stark shift can be made state independent in several ways. For example, one can choose $|q_{0}^{(n)}|^2=1/3$ to yield
\begin{equation}
\Delta E_{AC}=({\mathcal E}_1+{\mathcal E}_2) \frac{2 \bar{\omega}}{3(\bar{\omega}-\omega_1)}.
\end{equation}
In our experiments we chose the polarizations of one field, described by $\mathcal{E}_1$ to be linear ($|q_{0}^{(1)}|^2=1$) and the other field with circular polarization ($|q_{0}^{(2)}|^2=0$) and having twice the intensity of the linearly polarized field $\mathcal{E}_2=2 ~\mathcal{E}_1$. Then the AC-Stark shift is also state independent,
\begin{equation}
\Delta E_{AC}={\mathcal E}_1 \frac{2 \bar{\omega}}{(\bar{\omega}-\omega_1)}.
\end{equation}
In both cases, the AC-Stark shifts produce a global shift of all energy levels under the assumption of large laser detuning. The global shift will not affect the eigenstates of the Hamiltonian, therefore, the eigenstates are not changed when the light fields are turned on and off and the observed transition frequencies, that are proportional to energy differences, are not altered by these AC-Stark shifts. The validity of this approximation can be estimated by looking at the relative difference between energy denominators in Eq.(\ref{EquRamRab}). The contributing intermediate states are $J+1$ and $J-1$, therefore, the energy differs by  $\Delta E = \hbar R(2+4 J) \leq 26~ \hbar R$ for $J\leq 6$, $\Delta E\simeq h\times3.74$ THz for $J=6$. The energy difference between the ground and the vibrational ground state in the first excited electronic state is roughly $h\times430.8$ THz~\cite{abe2012}. Therefore, all energy denominators deviate by less than 1\%.\\
\\
In the experiment, the intensities of the $\sigma$ and $\pi$ beams are calibrated with the AC Stark shifts on the $|S, m_F = -1/2\rangle\leftrightarrow|D,m_F=-5/2\rangle$ transition. During the calibration, the ion order is switched such that the \Ca~ion is where the \CaH~would be in the spectroscopy experiments. The power in both beams is actively controlled to yield stable Stark shifts during experiments. AC Stark shifts of 200~kHz and 130~kHz for the $\sigma^-$- and $\pi$-polarized beams are used in the sideband pulses. We experimentally find minimal shifts in the transition frequencies inferred from sideband and carrier measurements at the ratio of 2:1.3 for the Stark shifts on \Ca.
\subsection{Trap parameters}
The ions are trapped in a linear Paul trap consisting of two wafers with segmented DC electrodes. Some details about the trap construction can be found in~\cite{leibfried05}. The secular frequencies for single \Ca~are \{$\nu_x$, $\nu_y$, $\nu_z$\}=\{4.567, 7.546, 3.000\} MHz. In the experiment, the order of the two-ion crystal is monitored and maintained the same throughout to improve the consistency of laser beam illumination and micro-motion compensation during the experiments.
\subsection{Sympathetic ground-state cooling}
Ground-state cooling before the spectroscopy experiments is achieved by an initial stage of Doppler cooling with the 397 nm $\pi$ beam (blue arrow perpendicular to the magnetic field in Fig.~1 of the main text), followed by EIT cooling with the 397 nm $\pi$ and $\sigma$ beams (blue arrow parallel with the magnetic field)~\cite{morigi00}, and resolved sideband cooling of the in-phase  (IP) and out-of-phase (OOP) modes in the $z$ direction and the lower frequency ($\sim 3.4$~MHz)  out-of-phase rocking (ROC) mode in the $x$ direction with the 729 nm beam (red arrow perpendicular to the B field). We use the $z$ OOP mode with low ($<1$ quanta/100 ms) heating rate for quantum-logic readout with low false-positive error rates. The $x$ ROC mode is cooled to the ground state to avoid adverse effects from the parametric coupling between that mode and the readout mode~\cite{roos08}. The $D_{5/2}$ and $D_{3/2}$ states are repumped by the 854 and 866 nm beams (pink arrows perpendicular to the B field in Fig. 1 of the main text).
\subsection{Pumping of molecular states}
The room temperature environment in our experiment equilibrates the \CaH~population with the background blackbody radiation (BBR) such that it is in the electronic and vibrational ground state with very high probability ($> 99$\%). The populations in the rotational levels are thermally distributed. For \CaH~in $B\sim 0.36$ mT, the frequencies of most $|J,m,\xi\rangle\rightarrow|J,m-1,\xi\rangle$, $\xi\in\{+,-\}$, transitions fall into narrow regions around -2 and -6 kHz, respectively. Pumping of the molecular population can be achieved by driving the OOP blue sideband (BSB) of those transitions followed by OOP mode sideband cooling pulses on \Ca~\cite{schmidt05}. The population is pumped toward the $-(J\pm\frac{1}{2})$ ends by ensuring that the frequency of the 1051 nm $\sigma^-$ beam is higher than that of the $\pi$ beam by the OOP mode frequency, $\nu_{\text{t}}\sim 5.164$~MHz, minus the frequency of the transition with $\Delta m = -1$ to be driven. Pumping of the spectroscopically resolved transitions $|J,-J+1/2,+\rangle \leftrightarrow |J,-J+3/2,-\rangle$ and $|J,-J-1/2,-\rangle \leftrightarrow |J,-J+1/2,-\rangle$ is interspersed to accumulate all of the molecular population in the states $|J,-J+1/2,-\rangle$ (see Fig. 2 of the main text for $J$=1,2).

The pumping pulses from the 1051 nm beams do not change the rotation quantum number $J$ and thus maintain the equilibrium population in each $J$ manifold dictated by the background blackbody radiation.
\subsection{Projective purification of imperfect $\text{Ca}^+$ state preparation}
\begin{figure}[!ht]
\vspace{-0.5 in}
\includegraphics[width=0.8\textwidth, right]{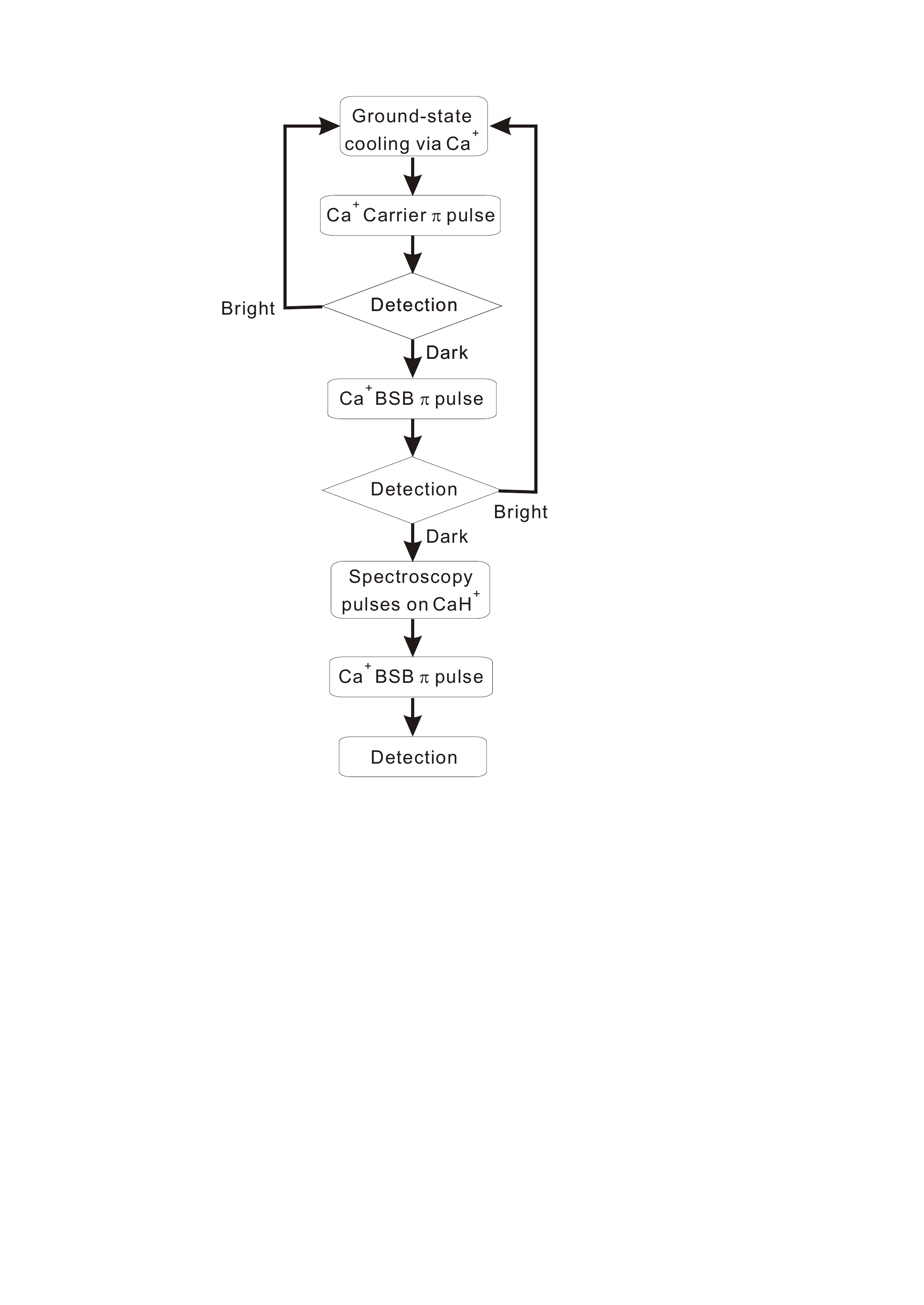}
\vspace{-3 in}
\caption[Flow chart.]{Flow chart of the purification stages.}
\label{FlowChart}
\end{figure}

To achieve a low false-positive error rate in the experiment, imperfect \Ca~state preparation and ground-state cooling are purified by exploiting the high detection fidelity in \Ca~state determination. With fluorescence count rates at $>2\times10^5$ and $<4\times10^3$ counts per second when the \Ca~is in the $|S\rangle$ and $|D\rangle$ states, respectively, the detection fidelity can be $>0.9999$~\cite{myerson08}. The target state, $|D\rangle|0\rangle$, is prepared via two purification stages. First, after ground-state cooling, a carrier $\pi$ pulse converts $|S\rangle$ to $|D\rangle$. If the \Ca~is found to be bright in the subsequent detection pulse, the state preparation starts over. Otherwise, the second purification stage follows, where a BSB $\pi$ pulse converts $|D\rangle|1\rangle$ to $|S\rangle|0\rangle$ but leaves $|D\rangle|0\rangle$ unaffected. The subsequent detection pulse would signal most of the population not in $|0\rangle$ with a bright outcome and state preparation would start again from the beginning. The two dark outcomes from the first two purification stages leave the population mostly in $|D\rangle|0\rangle$ and the ions are ready for a spectroscopy experiment. The spectroscopy pulse on the molecule will deposit a quantum of motion in the OOP mode if successful or leave that mode in the ground state otherwise. This is followed by a BSB $\pi$ pulse on the atomic ion, transforming $|D\rangle|1\rangle$ to $|S\rangle|0\rangle$. A bright outcome in the final detection pulse signals creation of a phonon in the OOP mode due to the spectroscopy experiments. A flow chart for the sequence is shown in. Fig.~\ref{FlowChart}. Using the purifying preparation sequence, false positive errors can be reduced to $\sim 0.5\%$ for spectroscopy experiments shorter than 1.5 ms. A similar sequence is also implemented when the $|D\rangle|1\rangle$ state is prepared for the spectroscopy experiments. Heating increases the false-positive error rates with increasing duration of the spectroscopy experiments.
\subsection{Experimental sequences for repetitive projective state preparation, Rabi spectroscopy, and Ramsey fringes}
%
Coherent spectra, Rabi flopping, and Ramsey fringes are obtained by repeating experimental sequences consisting of projective state preparation, molecular spectroscopy pulses, and state detection. The experiment control system attempts to keep track of the current state of the molecule. In the preparation stage, pumping is applied periodically if the rotational state of the molecule is unknown. Between pumping, projective state preparation for various $J$-manifold is attempted by applying Raman sideband $\pi$ pulses to drive the $|\mathcal{T}_J\rangle|0\rangle\rightarrow|J,-J-1/2,-\rangle|1\rangle$ transition and detecting the results. A positive event prompts subsequent projective state preparation attempts for the same $J$. We require at least two consecutive positive events signaling the transitions $|\mathcal{T}_J\rangle|0\rangle\leftrightarrow|J,-J-\frac{1}{2},-\rangle|1\rangle$ before the experiment control signals the successful preparation of the molecule in $|J,-J-\frac{1}{2},-\rangle$. The repetitive projection filters the false-positive events and thus achieves higher fidelity in projective state preparation. If four consecutive negative events are registered, the experiment control determines that the molecule is not in the corresponding manifold and switches to attempt state preparation on a target transition with different $J$. After the projective state preparation is heralded, the state of the \Ca~ion and the motional mode are prepared in $|D\rangle|0\rangle$. For the coherent spectra and Rabi flopping, we apply the carrier pulse with variable $\Delta\nu$ and pulse duration, respectively, to drive the $|J,-J-\frac{1}{2},-\rangle\leftrightarrow|\mathcal{T}_J\rangle$ transition. For the Ramsey fringes, two carrier $\pi/2$ pulses on the transition are applied, separated by a delay of duration $T$, with variable relative phase between the two pulses. We then detect whether the molecule has made the transition to $|\mathcal{T}_J\rangle=|J,-J+\frac{1}{2},-\rangle$. A sideband $\pi$ pulse on the molecular $|\mathcal{T}_J\rangle|0\rangle\leftrightarrow|J,-J-\frac{1}{2},-\rangle|1\rangle$ transition followed by a sideband pulse driving the $|D\rangle|1\rangle\rightarrow|S\rangle|0\rangle$ transition in \Ca~are applied. The state detection ends with a fluorescence detection with the outcome $|S\rangle$ signaling a positive molecular transition event. By repeating the sequence we can measure the transition probabilities.

\subsection{Comparison of calculated and experimental molecular transition parameters}
We presented spectroscopy experiments on the motional sidebands of the $|\mathcal{T}_J\rangle\leftrightarrow|J, -J-1/2,-\rangle$ transitions (see Fig.~3 in the main text), as well as carrier transitions (see Fig.~4, main text). Determination of energy differences based on the carrier transitions should reduce systematic uncertainties as compared with sideband transitions for two reasons: (i) Drifts of the secular frequency over time does not contribute to the uncertainties in the measured resonance frequencies. (ii) For a given Rabi rate, the light intensity on a carrier transition can be reduced by a factor of $\eta$, the Lamb-Dicke parameter, compared with a sideband transition. This significantly reduces the differential AC Stark shift induced by the probe laser beams. Table~\ref{tab_carr_freq} lists the experimental carrier transition frequencies for the rotational manifolds $J=1,2,\ldots,6$ at 1/8 of the light intensity used for the sideband transitions shown in Fig.~3 in the main text, as well as the predictions from our theoretical model. Experimental errors are purely statistical and theoretical uncertainties are based on the assumption that uncertainties in the parameters $c_{IJ}$ and $g$ add like random and independent variables. While the proximity of the experimental frequencies to the calculated frequencies supports our identification of the transitions, we are unable to perform a quantitative comparison, which would require careful characterization of systematic uncertainties in the experiment and of correlations between the theoretical uncertainties.

\begin{table}
\caption{Comparison between experiment and theory for spectroscopy on the carrier transitions. Listed are the observed and calculated center frequencies of the transitions $|\mathcal{T}_J\rangle\leftrightarrow|J, -J-1/2,-\rangle$. Experimental uncertainties are statistical, indicating the 68~\% confidence interval. Theoretical values are computed using the spin-rotation constants and g factors from Table~\ref{tab_theory}, the magnetic-field value B = 0.357~mT and taking into account off-resonant coupling of the levels $|\mathcal{T}_J\rangle$ and $|J, -J-1/2,-\rangle$ to other magnetic sublevels when driven by the Raman beams. Theoretical uncertainties are based on a 5~\% estimate for the relative
uncertainties of the constants, using Gaussian error propagation. It should be noted that, unlike in the case of statistical uncertainties, the theoretical uncertainties might be correlated and Gaussian error propagation is hence only an approximation.}
\begin{tabular}{lcc}
\\
\hline
\hline
\multicolumn{1}{c}{$J$}& \multicolumn{1}{c}{Experiment (kHz)} & \multicolumn{1}{c}{Theory (kHz)
}\\
\hline
\\ $1$        & 10.94(14)  & 10.73(44)
\\ $2$        & 13.55(8)  & 13.51(80)
\\ $3$        & 18.42(10)  & 18.90(1.27)
\\ $4$        & 24.90(15)  & 25.56(1.75)
\\ $5$        & 31.79(7)  & 32.85(2.20)
\\ $6$        & 39.51(5)  & 40.54(2.64)
\\
\hline
\hline
\end{tabular}
\label{tab_carr_freq}
\end{table}
We also compared the experimentally determined Rabi rates from the Rabi flopping curves in Fig.~4 in the main text with theory. For the probed transitions in the $J = 1$ and $J = 2$ manifolds, we find the Rabi rates $\Omega_{J=1} =
2\pi\times2.078(14)$~kHz and $\Omega_{J=2} = 2\pi\times1.804(12)$~kHz, respectively (uncertainties are statistical). The ratio of the two is 1.152(11), in reasonable agreement with the value 1.132 predicted by Eqs.~(\ref{EquRamFac}) and (\ref{EquSumAng3}).

\subsection{Data Availability}
Data are available from the authors upon request.

\end{document}